\documentclass[fleqn,twoside]{article}
\usepackage{espcrc2}

\usepackage{graphicx}
\usepackage[figuresright]{rotating}

\newcommand{\AmS}{{\protect\the\textfont2  A\kern-.1667em\lower.5ex\hbox{M}\kern-.125emS}}

\title{Measurement of \as\ in Radiative Hadronic Events at OPAL}

\author{Jochen Schieck \address{Max-Planck-Institut f\"ur Physik\\ F\"ohringer Ring 6, 80805 M\"unchen, Germany} }

\newcommand{\as}                {\ensuremath{\alpha_\mathrm{S}}}
\newcommand{\asmz}   {\ensuremath{\alpha_s(M_\mathrm{Z})}}
\newcommand{\epem}              {\ensuremath{\mathrm{e^+e^-}}}
\newcommand{\rs}                {\ensuremath{\sqrt{s}}}
\newcommand{\rsp}    		{\ensuremath{\sqrt{s'}}}
\newcommand{\Z}      {\ensuremath{\mathrm{Z}^0}}
\newcommand{\bm}[1]  {\mbox{\boldmath\ensuremath{#1}}}
\newcommand{\mh}     {\ensuremath{M_H}}
\newcommand{\thr}    {\ensuremath{1-T}}
\newcommand{\bt}     {\ensuremath{B_T}}
\newcommand{\bw}     {\ensuremath{B_W}}
\newcommand{\Mz}     {\ensuremath{M_{\mathrm{Z}}}}
\newcommand{\Jetset} {\texttt{JETSET}}
\newcommand{\Herwig} {\texttt{HERWIG}}
\newcommand{\Ariadne}{\texttt{ARIADNE}}

\newcommand{\lamqcd} {\ensuremath{\Lambda_{\overline{MS}}^{(5)}}}

\begin{document}

\begin{abstract}

Hadronic final states with a hard isolated photon are studied using
data taken at centre-of-mass energies around $\Mz$ with the OPAL
detector at LEP.  The strong coupling $\as$ is extracted by fitting
event shape variables for the reduced centre-of-mass energies ranging
 from 20~GeV to 80~GeV, and the energy dependence of
$\as$ is studied.  Combining all the values using different event shape
variables and energies gives:
\[
 \asmz=0.1176\pm 0.0012(\mathrm{stat.})^{+0.0093}_{-0.0085}(\mathrm{syst.}). 
\]
\vspace{1pc}
\end{abstract}

\maketitle

\section{Introduction}

The structure of QCD predicts a decrease of the strong coupling constant \as\
for high energy or equivalently for short distance reactions. At \epem\ collider
experiments the energy scale is provided by the center-of-mass energy \rs. 
The LEP collider covers energy ranges from 91 to 210 GeV~\cite{AsLEP1,AsLEP2-130GeV,AsLEP2}.
The change of \as\ at these energy ranges is relatively small compared to 
lower center-of-mass energies and therefore it is desirable to access \as\ at a lower \rs. \\
The radiation of high energy photons either through initial state radiation (ISR) or through 
quark bremsstrahlung (FSR) allows to access lower center-of-mass energies.
To make sure that the photon does not interfere with QCD processes 
the time scale of the photon radiation has to be smaller than the time scale
of the parton shower. With radiative multi-hadronic events 
(i.e. $\epem\rightarrow\mathrm{q}{\overline{\mathrm{q}}}\gamma$
a measurement of $\as$ is possible at a reduced centre-of-mass energy, $\rsp$.
At LEP1, isolated high energy photons observed in the detector are mainly photons
originating from FSR.
Here we report on a measurement of $\as$ from event shape variables of
the hadronic system in events with observed photons in the OPAL experiment~\cite{OpalDetector}.
The OPAL detector operated at the LEP $\epem$ collider at CERN.  
A more detailed description of the analysis can be found in~\cite{PN519}.

\section{Event Selection}

\subsection{Hadronic Event Selection}
\label{MHSelection}

This study is based on a sample of 3 million hadronic $\Z$ decays
selected as described in~\cite{LineShape} from the data accumulated
between 1992 and 1995 at centre-of-mass energy of 91.2 GeV.  We
required that the central detector and the electromagnetic calorimeter
were fully operational.

\subsection{Isolated Photon Selection}

\subsubsection*{Isolation Cuts}

A signal event is defined as an 
$\epem\rightarrow\mathrm{q}\mathrm{\overline{q}}$ event with an
initial or final state photon with energy greater than 10~GeV.
The selected photon candidate is required to be well contained
within the detector. In addition the candidate cluster is required 
to be well isolated from other clusters and tracks.
The angle of the candidate cluster with respect to the axis of any 
jet, $\alpha^{\mathrm{iso}}_{\mathrm{jet}}$, has to
to be larger than $25^\circ$.

\subsubsection*{Likelihood Photon Selection}

Isolated photon candidates are selected by using a likelihood
ratio method with four input variables. The first two variables are the same as described above,
two more variables are defined to reduce 
the background from clusters arising from the decay of neutral hadrons. 
The cluster shape fit variable $C$ determines the difference between
the measured energy deposit in a electromagnetic calorimeter and the
expected energy deposit from a simulated single photon. 
The second new variable is a measure of the distance between the 
electromagnetic calorimeter cluster and the associated 
presampler cluster.
A disagreement between data and Monte Carlo is seen for $C$ and
$\alpha^{\mathrm{iso}}_{\mathrm{jet}}$.  
It is related to the difficulty in predicting the
rate of isolated neutral hadrons in the Monte Carlo generators, as explained in
Section~\ref{BkgEstimation}. In this analysis, the rate of
isolated neutral hadrons used in the background subtraction
is estimated from data.

Hadronic events with hard isolated photon candidates are divided into
seven subsamples according to the photon energy for further \mbox{analysis.}
Table~\ref{SelectedEvents} shows the mean values of
$\rsp$($=2E_{\mathrm{beam}}\sqrt{1-E_{\gamma}/E_{\mathrm{beam}}}$),
and the number of data events.

\begin{table}[h]
\caption{The selected number of events and the mean value of $\sqrt{s\prime}$
for each $\sqrt{s\prime}$ subsample.}
\label{SelectedEvents}
\begin{tabular}{lll}
\hline
$E_{\gamma}$[GeV]  & Events  & $\sqrt{s\prime}_{Mean}$[GeV] \\
\hline
10-15 & 1200 & $78.0\pm1.7$ \\
15-20 & 764 & $71.7\pm1.8$ \\
20-25 & 511 & $65.0\pm2.1$ \\
25-30 & 418 & $57.6\pm6.7$ \\
30-35 & 383 & $48.9\pm2.6$ \\
10-15 & 303 & $38.5\pm3.4$ \\
10-15 & 248 & $24.3\pm5.3$ \\
\hline
\end{tabular}
\end{table}

\subsection{Background Estimation}
\label{BkgEstimation}

As mentioned in~\cite{PromptPhoton}, the \Jetset\ Monte Carlo fails to
reproduce the observed rate of isolated electromagnetic clusters, both
from isolated photons and from isolated $\pi^0$'s. 
Isolated neutral hadrons are the dominant source of background
for this analysis, and their rate has been
estimated directly from data using the following two methods. \\
The observed likelihood distributions in the data in bins of photon
energy were fitted with a linear combination of the Monte Carlo
distributions for signal and background events which pass the
isolation cuts and likelihood preselection requirements.  The overall
normalization of the Monte Carlo distribution is fixed to the number
of data events. The fit is a binned maximum likelihood with the
fraction of background events as a free parameter. \\
Alternatively the fraction of background from isolated
neutral hadrons was estimated from the
rates of isolated charged hadrons. When isospin
symmetry is assumed, the rates of neutral pions, neutral kaons and neutrons
can be estimated from the rates of charged pions, charged kaons and protons,
respectively.
According to \Jetset\ tuned with OPAL data, the deviation due to
isospin violating decay is predicted to be 10\% for pions and 5\% 
for kaons and protons. This
is assigned as a systematic uncertainty for the
isolated tracks method and combined with the statistical uncertainty.
Consistent results are obtained from the two methods within the
errors.   \\

The non-radiative multihadron background varies
between $15.7\pm3.5 \%$ and $4.4\pm0.8 \%$. The contamination
from $\tau \tau$ background is about $1\%$.
The contribution of two photon processes is less than 0.01\% in all subsamples.

\section{Global Event Shape Variables}

The determination of $\as$ is based on measurements of
event shape variables, which are calculated from particles with momenta $p_i$
in an event. In this analysis the following event shape variables are used:
Thrust \bm{T}~\cite{thrust}, Heavy Jet Mass \bm{M_H}~\cite{hjm}
and the Jet Broadening variables \bm{B_T} and \bm{B_W}~\cite{jbv}.

The event shape variables are calculated from tracks and
electromagnetic clusters excluding the isolated photon candidate.
Contribution of electromagnetic clusters originating from charged particles
are removed by the method described in~\cite{CharginoMT}.\\
Since these variables are defined in the centre-of-mass frame of the colliding beams, the
hadronic system is boosted back into the centre-of-mass frame
of the hadrons. The Lorentz boost is determined from the energy and
angle of the photon candidate. 
The contribution from non-radiative
hadronic events is removed statistically by subtracting the Monte
Carlo distribution scaled by the fraction of non-radiative hadronic 
events and $\tau$ pair events listed in Table~\ref{SelectedEvents}.  The
effects of the experimental resolution and acceptance
are unfolded using Monte Carlo samples with full detector
simulation (detector correction).  
We refer to the distributions after applying these
corrections as data corrected to the hadron level.  \\

\section{Measurement of \as\ from Event Shape Distribution}

The measurement of $\as$ is performed by fitting perturbative QCD
predictions to the event shape distributions corrected to the
hadron level for $(\thr)$, $\mh$~\cite{CalcThrustMH}, $\bt$ and
$\bw$~\cite{CalcBTBW}. The {\cal O($\as^2$)} and NLLA calculations are
combined with the ln($R$) matching scheme. 
The hadronization correction is applied to the cumulative theoretical
calculation to conserve normalization as in our previous analysis at
centre-of-mass energies of 130~GeV and 136~GeV and
above~\cite{AsLEP2-130GeV,AsLEP2}.  
\Jetset, \Herwig\ and \Ariadne\ are used for this
hadronization correction and \Jetset\ is chosen for the central results.\\
The fit to the event shape variables uses 
a least $\chi^2$ method with $\as(Q)$ treated as a free parameter.
When the total number of events is small, the differences in statistical
error between bins with a larger or smaller number of events than the
theoretical prediction is not negligible.  To deal with fluctuations of
the error, the value of the fitted theoretical distribution is used instead of
the number of events in each bin of the data distribution.
The statistical uncertainty is estimated from fit results derived from
100 Monte Carlo subsamples with same number of events as selected data
events.
The background subtraction, detector and hadronization corrections are
required to be small and uniform in the region used for the fit.

\subsection{Systematic Uncertainties}
\label{SystUncertainties}
The total systematic uncertainties where estimated by adding 
experimental, hadronization and theoretical uncertainties in quadrature.
The largest contributions originate from the choice of the hadronization
model and the variation of the renormalization scale.

\subsection{Combination of \as\ Results}

\label{Combination}
\begin{figure}[h]
\includegraphics[scale=0.25]{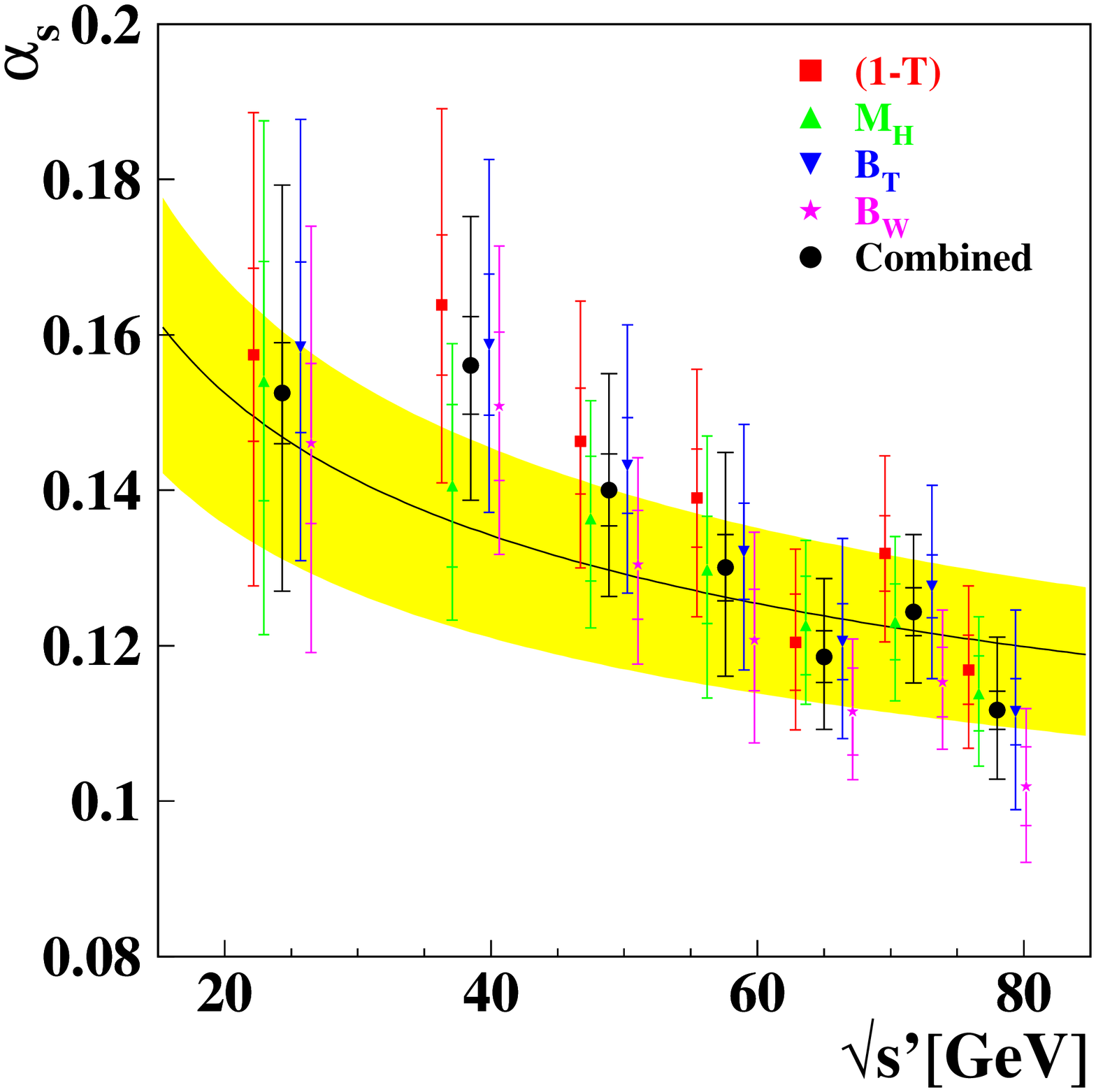}
\caption{Energy dependence of \as\ for all subsamples.}
\label{AsEdep} 
\end{figure}

The values of $\as$ obtained by fits to event shape variables at each
energy are used to study the energy dependence of $\as$, and to
obtain an overall combined result for $\asmz$. \\
Values of $\as$ for each event shape variable and for all event
shape variables combined 
are fitted to the solution of the renormalization group equation at
NNLO described in~\cite{PDG}.  $\lamqcd$ is
treated as a free parameter in the fit. When combining all event shape
variables, the statistical correlations between results from different
variables are taken into account.
The systematic uncertainties for $\lamqcd$
from each variable are obtained by the procedure described in
Section~\ref{SystUncertainties}.
The value of $\lamqcd$ derived from the values of $\as$ combining the event shape variables is
\begin{equation}
 \lamqcd = 0.2027\pm 0.0141(\mathrm{stat.})^{+0.1130}_{-0.0939}(\mathrm{syst.})\ \mathrm{GeV}
\end{equation}
with $\chi^2/\mathrm{d.o.f}=33.1/6$.
The systematic uncertainty has contributions from experimental effects 
($\pm0.0417$), hadronization effects ($\pm0.0668$) and variations of 
renormalization scale ($+0.0810,-0.0512$).
This value and its total errors correspond to 
$\asmz = 0.1175^{+0.0085}_{-0.0102}$ in NNLO.
QCD prediction of $\as$ with $\lamqcd$ obtained by the fitting is shown in 
Figure \ref{AsEdep}. 
The  $\chi^2$  value is dominated by the subsample with $\rsp$ of 38.5GeV,
the othe subsamples agree well with the prediction.

All values of $\as$ are propagated to the energy scale of $\Mz$ using the
equation for the evolution of $\as$ quoted in~\cite{PDG}. 
The combined values and values for individual event shape variables 
for all $\rsp$ subsamples are
combined into one value for all systematic variations.  Since there is
no statistical correlation between values for different $\rsp$
subsamples, a weighted mean is calculated using statistical uncertainties only.
The systematic uncertainties on the combined values are obtained by
the procedure described in Section~\ref{SystUncertainties}.  
These other values are consistent with each
other and in good agreement with the present result.  As pointed out in
the analysis with non-radiative events at LEP1~\cite{AsLEP1} and
LEP2~\cite{AsLEP2}, the $\as$ value obtained by fitting $\bw$ is lower
than for the other three variables
The combined value of $\asmz$ for all $\rsp$ subsamples and all event shape
variables is
\begin{equation}
 \asmz=0.1176\pm 0.0012(\mathrm{stat.})^{+0.0093}_{-0.0085}(\mathrm{syst.}). 
\end{equation}

The systematic uncertainty has contributions from experimental effects 
($\pm0.0034$), hadronization effects ($\pm0.0061$) and variations of 
renormalization scale ($+0.0062,-0.0049$).
This value is consistent with the result from the analysis using
non-radiative events in LEP1 data, $\asmz=0.120\pm 0.006$.  More
event shape variables are fitted 
for the non-radiative events.  If the variables are
restricted to the set used in the present analysis,
the combined value from the non-radiative events is
$\asmz=0.1155^{+0.0071}_{-0.0060}$.

\section{Summary}

The strong coupling $\as$ has been measured at reduced centre-of-mass
energies, $\rsp$, ranging from 20~GeV to 80~GeV using the event shape
of the hadronic system in radiative hadronic events.  Four event shape
variables, $\thr$, $\mh$, $\bt$ and $\bw$ of the hadronic system boosted
into the centre-of-mass frame are fitted
with {\cal O($\as^2$)} and NLLA QCD predictions 
and values of $\as$ are obtained for several values of $\rsp$.  We obtained the
fundamental constant of QCD, $\lamqcd$, from
the energy dependence of these values of $\as$.  Values at each $\rsp$ are
evolved to $\mu=\Mz$ and combined for each event shape variable. The
combined value from all event shape variables and $\rsp$ values is 
 $\asmz=0.1176\pm 0.0012(\mathrm{stat.})^{+0.0093}_{-0.0085}(\mathrm{syst.})$.

This agrees with the previous OPAL analysis with non-radiative LEP1 data
and the world average PDG value.  Within errors, 
QCD is consistent with our data sample of events with isolated FSR.


\begin{thebibliography}{9}
\bibitem{AsLEP1}
P.~D.~Acton {\it et al.} 
Z.\ Phys.\ C {\bf 59} (1993) 1.

\bibitem{AsLEP2-130GeV}
G.~Alexander {\it et al.} 
Z.\ Phys.\ C {\bf 72} (1996) 191.
\bibitem{AsLEP2}
K.~Ackerstaff {\it et al.}
Z.\ Phys.\ C {\bf 75} (1997) 193;
G.~Abbiendi {\it et al.} 
Eur.\ Phys.\ J.\ C {\bf 16} (2000) 185.

\bibitem{OpalDetector} K.~Ahmet {\it et al.}
Nucl.\ Instrum.\ Meth.\ A {\bf 305} (1991) 275.

\bibitem{PN519} K.~Ahmet {\it et al.} 
OPAL Physics Note PN519 (unpublished)

\bibitem{LineShape}
G.~Alexander {\it et al.}  [OPAL Collaboration],
Z.\ Phys.\ C {\bf 52} (1991) 175.

\bibitem{PromptPhoton} 
K.~Ackerstaff {\it et al.}
Eur.\ Phys.\ J. \ C {\bf 5} (1998) 411. \\
O.~Adriani {\it et al.}
Phys.\ Lett.\ B {\bf 292} (1992) 472.

\bibitem{thrust}
S.~Brandt {\it et al.}, Phys.\ Lett. {\bf 12}(1964) 57; \\
E.~Farhi, Phys.\ Rev.\ Lett. {\bf 39}(1977) 1587.

\bibitem{hjm}
T.~Chandramohan and L.Clavelli, Nucl.\ Phys. {\bf B184} (1981) 365; \\
A.~Peterson {\it et al.}, Phys. \ Rev. {\bf D37} (1988)1; \\
W. Braunschweig {\it et al.}, Z. \ Phys. {\bf C45} (1989) 11. 

\bibitem{jbv}
S.~Catani {\it et al.}, Phys. \ Lett. {\bf B295} (1992) 269.

\bibitem{CharginoMT}
K.~Ackerstaff {\it et al.}  [OPAL Collaboration],
Eur.\ Phys.\ J.\ C {\bf 2} (1998) 213.

\bibitem{CalcThrustMH}
S.~Catani, L.~Trentadue, G.~Turnock and B.~R.~Webber,
Nucl.\ Phys.\ B {\bf 407} (1993) 3.

\bibitem{CalcBTBW}
Y.~L.~Dokshitzer, A.~Lucenti, G.~Marchesini and G.~P.~Salam,
JHEP {\bf 9801} (1998) 011.



\bibitem{PDG}
K.~Hagiwara {\it et al.}
Phys.\ Rev.\ D {\bf 66} (2002) 010001.


\end{thebibliography}
\end{document}